\documentclass[12pt]{article}

\usepackage{amsfonts,amssymb}
\newcommand{\dd}{ \hbox{d} }

\begin{document}

\title{\bf The Complete Black Brane \\Solutions in $D$-Dimensional 
\\
Coupled Gravity System}
\author{Bihn Zhou\thanks{E-mail: zhoub@itp.ac.cn}\, and 
Chuan-Jie 
Zhu\thanks{
E-mail: zhucj@itp.ac.cn} \\
Institute of Theoretical Physics, Chinese Academy of\\
Sciences, P. O. Box 2735, Beijing 100080, P. R. China}

\maketitle

\begin{abstract}
In this paper, we use only the equation of motion for an interacting system 
of gravity,
dilaton and antisymmetric tensor to study the black brane solutions.  By 
making use
of the property of Schwarzian derivative, we obtain the complete solution 
of 
this system of  equations. For some special values we obtain the well-known 
BPS brane  and black brane solutions. 
\end{abstract}

\newpage
\section{Introduction}
Superstring theory \cite{GSW, Joe} is the leading candidate for a theory 
unifying  all matter and
forces (including gravity, in particular).  Unfortunately, there are five such 
consistent 
theories, namely, $SO(32)$ type I , type IIA, type IIB, $E_8 \otimes E_8$ 
heterotic
and $Spin(32)/Z_2$ heterotic theories. This richness is an embarrassment 
for pure
theorists.  One hope is that all these five superstring theories are just
different
solutions of an underlying theory. In the last few years this hope turns out to 
be true.
The underlying theory is the so called M theory \cite{Hull, Wittena, Schwarz}. 
One  formulation of M theory is given in terms of Matrix theory \cite{BFSS}. 
Nevertheless this formulation is not background independent. It is difficult to 
use it to discuss non-perturbative problems and still we should rely on the 
study of
BPS states. A thorough study of BPS states in any theory is a must for an
understanding of  non-perturbative phenomena.  

In superstring theory and M theory, there is a plethora of BPS states. These 
BPS
states have the special property of preserving some supersymmetry.  In a low 
energy
limit they are special solutions of the low energy supergravity theory with
Poincar\'e invariance. However there exists also some other $p$-branes which
 don't
preserve any supersymmetry and they are also no longer Poincar\'e invariant. 
In finding these (soliton) solutions one either resort to supersymmetry or make
some simple plausible assumptions. There is no definite reasoning that the
so obtained solution is unique. The purpose of this paper is to fill this gap.

It is also interesting to study soliton solutions for its own sake without using
any supersymmetric argument. Quite recently there are some interests 
\cite{Klebanov,
Minahan, Ferretti, Bergman} in studying
branes in type 0 string theories which have no fermions and no supersymmetry
 in
10 dimensions \cite{Joe}.
In these string theories, we can't use supersymmetric arguments.
So it is important to push other symmtric arguments to their limits.

In this paper we will study the coupled system of gravity ($g_{MN}$), dilaton
($\phi$) and anti-symmetric tensor ($A_{M_1M_2\cdots M_{n-1}}$)
in any dimensions. In a previous letter \cite{ZhouZhub}
we announced the complete solution
of this system. Here we will give more details of the derivation of the complete
solution and also extend the solution to include black branes. 
After making a quasi-Poincar\'e invariant ansatz for the metric
and either electric or magnetic ansatz for the  anti-symmetric tensor, we derive
 all the
equations of motion in some details.  A system of six ordinary differential
 equations
are obtained for five unknown functions (of one variable). 
After making some changes of  the 
unknown functions we solved four equations explicity.  As we shown in 
\cite{ZhouZhu},
these  remaining 
two equations (for one unknown functions) are mutually  compatible and a 
unique
solution can always be obtained with appropriate boundary conditions.  
The last equation can also be solved in the most generic case by exploiating 
some properties of Schwarzian derivative. The complete solutions are given 
explicitly.
By requiring that the solution approaches the flat space-time for $r \to
\infty$, we found that there remains  four free parameters.  For special
values of these four
parameters we got
the well-known BPS brane solution and black brane solution \cite{Strominger, 
Duffa, Duffb}.
We will not discuss the physical property of these solutions \cite{Zhou}.
For previous studies
of soliton and brane solutions in supergravity and string theories, see for 
example
the reviews \cite{Duff, Stelle, Argurio, Abers, Fre}. We note that some 
solutions are also
generated via sigma-models in \cite{Rytchkov}. These solutions are only 
special cases
of our complete solutions. In fact there is a large number of papers dealing
 with the
brane solution in one way or another. We apologize in advance that this paper 
is not a review and we can't cite all of them because the purpose of this
paper is to obtain the most
complete solution which wasn't achieved  before. 

\section{The Equations of Motion and the Ansatz}

Our starting point is the following action for the coupled system of gravity
$g_{MN}$, dilaton $\phi$ and anti-symmetric tensor
$A_{M_1\cdots M_{n-1}}$:
\begin{equation}
I = \int \dd^D x \sqrt{- g} \,
\left( R - { 1\over 2} g^{MN} \partial_M \phi \partial_N \phi -
{1 \over 2\cdot n! } e^{a\,\phi} F^2 \right),
\label{Action}
\end{equation}
where $a$ is a constant and $F$ is the field strength: $F=\dd A$.

The equations of motion can be easily derived from the above action 
(\ref{Action}):
\begin{eqnarray}
  & & R_{MN} = \frac{1}{2} \partial_M\phi\partial_N\phi + S_{MN},
\label{Req} \\
  & & {1\over \sqrt{-g} }\partial_{M_1} ( \sqrt{-g} \,
e^{a\phi} F^{M_1 M_2 \cdots M_n})  = 0,
\label{Feq} \\
  & & {1\over \sqrt{-g} } \partial_M( \sqrt{-g} g^{MN} \partial_N\phi)
= {a\over 2 n!}\, e^{a\phi}\, F^2,
\label{phieq}
\end{eqnarray}
where
\begin{equation}
  S_{MN} = {1\over 2(n-1)!} e^{a\phi}\, \left(F_{M M_2 \cdots M_n}
	F_N^{\ M_2 \cdots M_n} -{n-1\over n(D-2)} F^2 g_{MN}\right).
\end{equation}
Of course it is impossible to solve the above system of equations in their full
generality. To get some meaningful solution we will make some assumptions
by using symmetric arguments.

Our ansatz for a $p$-dimensional black brane is as follows:
\begin{equation}
\dd s^2 = - e^{2 A_0(r)} \dd t^2 +
\sum_{\alpha=1}^p e^{2 A_{\alpha}(r)} (\dd x^{\alpha})^2+
e^{ 2 B(r)} \dd r^2 +
e^{ 2 C(r)} \dd \Omega_{\tilde{d}+1}^2, 
\label{ansatz}
\end{equation}
where $\dd \Omega_{\tilde{d}+1}$ is the square of the line element on
the unit $\tilde{d}+1$
sphere which can be written as follows:
\begin{equation}
\dd \Omega_{\tilde{d}+1}^2 = \dd \theta_1^2 + \sin^2 \theta_1 \dd \theta_2^2
+ \cdots +
(\sin\theta_1 \cdots \sin\theta_{\tilde{d}})^2 \dd\theta^2_{\tilde{d}+1}.
\end{equation}
The brane is extended in the directions $(t, x^{\alpha})$.  When all 
$A_{\alpha}$ 
(including
$A_0$) are equal, 
the brane is Poinc\'are invariant in the space-time spanned by 
$(t, x^{\alpha})$. 
 We will 
call the ansatz (\ref{ansatz}) quasi-Poincare invariant when all 
$A_{\alpha}$ are equal except $A_0$.

For the anti-symmetric tensor $A$, we have 2 different choices. The first
 choice is
the electric case  and  we take the following form for $A$:
\begin{equation}
A = \pm e^{\Lambda(r)}\, \dd x^0\wedge \dd x^1 \wedge \cdots \wedge
 \dd x^p.
\end{equation}
The second choice is the magnetic case and  we take the following form for
the dual potential $\tilde{A}$:
\begin{equation}
\tilde{A} = \pm e^{\Lambda(r)}\, \dd x^0\wedge \dd x^1 \wedge \cdots \wedge
 \dd x^p.
\label{ansatzb}
\end{equation}
We note that the relation between the antisymmetric tensor field strength $F$
and its dual field strength $\tilde{F}$ ($\equiv \dd \tilde{A}$) is:
\begin{equation}
F^{M_1\cdots M_n} ={ 1\over \sqrt{ -g } } \, e^{ -a \phi} \, { 1\over (D-n)! }
 \,
\epsilon^{M_1 \cdots M_n N_1 \cdots N_{D-n} } \, \tilde{F}_{
N_1 \cdots N_{D-n} } .
\end{equation}
By using this relation the ansatz (\ref{ansatzb}) transform to an ansatz
for $F$:
\begin{equation}
F = \pm \Lambda' \hbox{exp}\left[ \Lambda - a \phi -\sum_{\alpha=0}^p 
A_{\alpha}
-B + (\tilde{d} +1 )\, C\right] \, \omega_{\tilde{d}+1},
\end{equation}
where $\omega_{\tilde{d}+1}$ is the volume form of the sphere
$S^{\tilde{d}+1}$ with unit radius.
As it is well-known from duality, the equation of motion (\ref{Feq}) becomes
 the
Bianchi identity for $\tilde{F}$ which is satisfied automatically. Nevertheless 
the
Bianchih identity for $F$ becomes the equation of motion for $\tilde{F}$  
which is
given as follows:
\begin{equation}
{1\over \sqrt{-g} }\partial_{N_1} ( \sqrt{-g} \,
e^{-a\phi} \tilde{F}^{N_1 N_2 \cdots N_{D-n}})  = 0.
\label{Feqa}
\end{equation}
From now on we will discuss the electric case only.

Denoting the coordinate index as
$(M) = (t, \alpha, r, i)$ and tangent index as $(A)=(\bar{t},\bar{\alpha},
\bar{r}, 
\bar{i})$,
i.e. a bar over coordinate index, we introduce the following moving frame:
\begin{eqnarray}
e^{\bar{t}} & = & e^{A(r)} \, \dd t = - e_{\bar{t}} , \\
e^{\bar{\alpha}} & = & e^{A_{\alpha}(r) } \,  \dd x^{\alpha} =
 e_{\bar{\alpha}}, \\
e^{\bar{r}} & = & e^{ B(r)} \, \dd r = e_{\bar{r}}, \\
e^{\bar{i}} & = & e^{ C(r)} \, \sin\theta_1 \cdots \sin\theta_{i-1}\dd 
\theta_{i}  =
 e_{\bar{i}} .
\end{eqnarray}
From \cite{ZhouZhu} we have the following results for the Riemann 
curvature:
\begin{equation}
R^{AB} = f^{AB} \, e^{A}\wedge e^{B},
\end{equation}
where there is no summation ove $A$ and $B$. $f^{AB}$ can be chosen to 
be symmetric
 and is 0
when $A=B$. By definition
\begin{equation}
R^{AB} = {1\over 2} \, R^{AB}_{MN} \, \dd x^M\wedge \dd x^N,
\end{equation}
we have
\begin{equation}
R^{AB}_{MN} = f^{AB}(e^A_M\, e^B_N - e^A_N \, e^B_M),
\end{equation}
and
\begin{equation}
R^A_M = R^{AB}_{MN}\, e^N_B = \sum_B f^{AB} \, e_M^A \equiv f^A
 \, e^A_M,
\end{equation}
where we have defined $f^A$ as
\begin{equation}
f^A=\sum_B f^{AB},
\end{equation}
which can be computed to give the following results:
\begin{eqnarray}
f^{\bar{t}} & = & -e^{-2B}\Big( A_0'' + A_0'(\sum_{\alpha=0}^p A_{
\alpha}' - B'
+ (\tilde{d}+1)C')\Big),
\\
f^{\bar{\alpha}} & = & - e^{-2 B}
\Big( A_{\alpha}'' + A_{\alpha}'( \sum_{\beta=0}^p A_{\beta}' - B'
+ (\tilde{d}+1)C')\Big),
\\
f^{\bar{r}} & =& -e^{-2B} \Big(  \sum_{\alpha=0}^p A_{\alpha}''
+ (\tilde{d}+1)C''  +
\sum_{\alpha=0}^p (A_{\alpha}')^2 \nonumber \\
&& + (\tilde{d}+1)(C')^2 -B'(  \sum_{\alpha=0}^p  A_{\alpha}' + (\tilde{d}
+1)C')
\Big) ,
\\
f^{\bar{i}}  & = & - e^{-2B}\, \Big( C'' + C'( \sum_{\alpha=0}^p A_{\alpha}'
-B'+ (\tilde{d}+1)C')
\nonumber \\
& & - \tilde{d}\, e^{-2C +2B}\Big).
\end{eqnarray}
Notice that the metric in (\ref{ansatz}) is a diagonal metric we have then
\begin{eqnarray}
R_{MN} & = & R^A_M \, e_{AN} = f^A\delta^A_M \, g_{MN},
\\
R_{MN}\dd x^M \otimes \dd x^N & = &
- f^{\bar{t}} \, e^{\bar{t}} \otimes e^{\bar{t}} +
\sum_{\alpha=1}^p f^{\bar{\alpha}} \, e^{\bar{\alpha}} \otimes e^{\bar{
\alpha}}
\nonumber \\
& & +  f^{\bar{r}} \, e^{\bar{r}} \otimes e^{\bar{r}} + f^{\bar{i}} \dd
\Omega_{\tilde{d}+1},
\end{eqnarray}

To obtain the equation of motion from eqs. (\ref{Req}) to (\ref{phieq})
we also need to know some expressions
involving $F$. We have (for elementary solution)
\begin{eqnarray}
A & = & \pm e^{\Lambda(r)} \, \dd x^0 \wedge \dd x^1 \cdots \wedge \dd x^{
n-2},
\\
F & = & \dd A = \pm \Lambda'\,e^{\Lambda(r)} \, \dd r \wedge \dd x^0 \wedge 
\dd x^1 
\cdots \wedge \dd x^{n-2},
\\
F_{MN} & \equiv  & F_{MM_2\cdots M_n} {F_N}^{M_2\cdots M_n} , \\
F_{rr} & = & -(n-1)! \, (\Lambda'\,e^{\Lambda(r)} )^2 \, e^{ -2\sum_{\alpha=
0}^p A_{
\alpha} } , \\
F_{\mu\nu} & = & -g_{\mu\nu} (n-1)! \, (\Lambda'\,e^{\Lambda(r)} )^2 \, e^{
 -2
\sum_{\alpha=0}^p A_{\alpha} -2B}, \\
F_{MN} & = & 0, \qquad \hbox{for the rest cases,}
\end{eqnarray}
and
\begin{equation}
F^2 = g^{MN}\, F_{MN} = - n!\, (\Lambda'\,e^{\Lambda(r)} )^2 \, e^{ -2
\sum_{\alpha=0}^p A_{\alpha}  -2B},
\end{equation}

The equation of motion for $g^{MN}$ gives the following equations:
\begin{eqnarray}
 f^{\bar{r}} g_{rr} & = &{1\over 2}(\partial_r \phi)^2 + { 1\over 2\cdot (n-1)!
}\,
 e^{a \,\phi} \left(F_{rr} - {(n-1)\over n(D-2) }\, F^2 \, e^{2B} \right),
\\
 f^{\bar{\mu}}\, e^{2A}\eta_{\mu\nu} & = &
{ 1\over 2\cdot (n-1)!}\,
 e^{a \,\phi} \left(F_{\mu\nu} - {(n-1)\over n(D-2) }\, F^2 \, e^{2A}\eta_{\mu
\nu} 
\right),
\\
 f^{\bar{i}} g_{ij} & =& { 1\over 2\cdot (n-1)!}\,
 e^{a \,\phi} \left(F_{ij} - {(n-1)\over n(D-2) }\, F^2 \, g_{ij} \right),
\end{eqnarray}
for the $(rr)$, $(\mu\nu)$ and $(ij)$ components respectively. Substituting all
$f^{A}$ and $F_{MN}$ into the above equations and setting $C =B + \ln r$,
$A_0 = A + {1\over 2}\, \ln f $ and $A_{\alpha} = A$ ($\alpha = 1, \cdots, p$),
we obtain the following four equations:
\begin{eqnarray}
& & A'' + d  (A')^2 + \tilde{d}\, A'B' + { \tilde{d}+1\over r } \, A'
+ {1\over 2} \, (\ln f)'\, A' =  { \tilde{d}\over 2 (D-2) } \, S^2 , 
\label{lasta}\\
& & A'' + d  (A')^2 + \tilde{d}\, A'B' + { \tilde{d}+1\over r } \, A' 
+ {1\over 2} \, (\ln f)''  + {1\over 2} \, (\ln f)' 
\nonumber \\
& & \qquad \times \left( (d+1) A' +{1\over 2} \, (\ln f)' + \tilde{d}
\, B' + { \tilde{d} + 1 \over r} \right) =  
{ \tilde{d}\over 2 (D-2) } \, S^2 , 
\\
& & B'' + d  A'B' + {d \over r} \,A' +\tilde{d} (B')^2  +{1\over 2} (\ln f)' 
\left( B' + { 1\over r}\right) 
\nonumber \\
& & \hskip 3cm + { 2 \tilde{d} + 1 \over r}\, B' = - { 1\over 2}\, 
{d \over D-2}\, S^2, 
\\
& & d A'' + (\tilde{d} + 1)B'' + d (A')^2 + { \tilde{d} + 1 \over r}\, B' 
- d A'B' + {1\over 2} \, (\ln f)''   \nonumber \\
& & \qquad + {1\over 2} \, ( 2 A' -B') +
{1\over 4} \, ( (\ln f)')^2  + { 1\over 2} \, (\phi')^2 = 
{1\over 2}\, {\tilde{d} \over D-2}\, S^2,
\label{lastb}
\end{eqnarray}
where $d=p+1=n-1$ and
\begin{equation}
  S= {\Lambda' \over f^{1\over 2} } \, \ e^{\frac{1}{2}a\phi+\Lambda- d A}.
\end{equation}
The equation of motion for $\phi$ is
\begin{equation}
\phi'' + \left( d A' + \tilde{d}B' + {\tilde{d} + 1 \over r} 
+ {1\over 2} \, (\ln f)' 
\right) 
\phi' = - { a\over 2} \, S^2 ,
\label{lastc}
\end{equation}
while the equation of motion for $F$ is
\begin{equation}
  \left( {\Lambda' \over f^{1\over 2} } \ e^{\Lambda+a\phi-d A+\tilde{d} B }
\, r^{\tilde{d}+1}
  \right)'=0.
\label{lastd}
\end{equation}
These six equations, eqs. (\ref{lasta})-(\ref{lastb}), (\ref{lastc}) and
(\ref{lastd}), consist of the complete system of equations of motion for
five unknow functions: $f(r)$, $A(r)$, $B(r)$, $\phi(r)$ and $\Lambda(r)$.
We will solve  these equations completely in the next three sections.

\section{A First Try at the Solution}

In this section we will try our best to solve the above system of equations by
elementary means. First it is easy
to integrate eq. (\ref{lastd}) to get
\begin{equation}
{\Lambda' \over f^{1\over 2}(r) } \, e^{\Lambda(r) + a \phi(r) - d A(r) + 
\tilde{d} B(r) }
\, r^{\tilde{d}+1} = C_0,
\label{czero}
\end{equation}
where $C_0$ is a constant of integration. If we know the other
four functions
$f(r)$, $A(r)$, $B(r)$ and $\phi(r)$, this equation can be easily
integrated to give $\Lambda(r)$:
\begin{equation}
e^{\Lambda(r)} = C_0\,  \int^r \dd r \, { f^{1\over 2}(r) \,  e^{ - a \phi(r)
+  d A(r) - \tilde{d} B(r)} \over  r^{\tilde{d}+1} } .
\end{equation}
By using eq. (\ref{czero}), $S$ can be written as  follows:
\begin{equation}
S(r) = C_0 \, { e^{ - {a \over 2 }\, \phi(r) - \tilde{d}\, B(r) } \over
r^{\tilde{d} +1} } .
\label{Sczero}
\end{equation}

In order to solve the other equations
we make a change of functions from $f(r)$, $A(r)$, $B(r)$ and
$\phi(r)$ to $\xi(r)$, $\beta(r)$, $\eta(r)$ and $Y(r)$:
\begin{eqnarray}
\xi(r) & = &  d A(r) + \tilde{d} B(r) +{1\over 2} \, \ln f(r) ,
\\
\beta(r) &=& d A(r) + \tilde{d} B(r) -{1\over 2} \, \ln f(r) ,
\\
\eta(r) & = & \phi(r) +   a\, \big( A(r)-B(r) \big),
\\
Y(r) & = & A(r)-B(r).
\end{eqnarray}
The equations  are then changed to
\begin{eqnarray}
& & \xi''+(\xi')^2+\frac{2\tilde{d}+1}{r} \, \xi' =0,
\label{xieq}  \\
& & \beta'' + \left( \xi'+\frac{\tilde{d}+1}{r} \right) \, \beta'
+ \frac{\tilde{d}}{r} \, \xi' = 0,
\label{betaeq} \\
& & \eta''+\left( \xi'+\frac{\tilde{d}+1}{r}\right)\, \eta'
  -\frac{ a}{r}\, \xi' =0,
\label{etaeq}  \\
& & Y''-\frac{\Delta}{2}(Y')^2+\left[\frac{\tilde{d} -d }{2(D-2)}\, (\xi +
\beta)'
  +\frac{\tilde{d}+1}{r}+ a\, \eta' -{1\over 2} \, (\xi-\beta)' \right]Y'
\nonumber \\
&  & \hskip 2cm  -\frac{1}{2}(\eta')^2-\xi''+\frac{(\xi'+\beta')^2}{4(D-2)}
  -{1\over 4}\, ( \xi'-\beta')^2  =0,
\label{Yeq}  \\
&  & Y''+\left(\xi'+\frac{\tilde{d}+1}{r} \right) \,Y' -\frac{\xi'}{r}
= \frac{1}{2}\,S^2,
\label{newSeq}
\end{eqnarray}
where
\begin{equation}
\Delta = \frac{2\,d\, \tilde{d}}{D-2} + a^2.
\end{equation}

The general solutions  for $\xi$, $\beta$ and $\eta$  can be obtained
easily from eqs. (\ref{xieq}), (\ref{betaeq})
and (\ref{etaeq}) and we have
\begin{eqnarray}
\xi  & = &\ln \left| C_1 + C_2 r^{-2\tilde{d}} \right|,
\label{sumdat} \\
\eta'  & = & \frac{ 2C_2   a + C_3 r^{\tilde{d}} }
  {r(C_2 + C_1 r^{2\tilde{d}})},
\label{phfdat} \\
\beta' &=& \frac{ - 2C_2 \, \tilde{d}  + C_4 r^{\tilde{d}} }
  {r(C_2 + C_1 r^{2\tilde{d}})},
\end{eqnarray}
where $C_i$'s are constants of integration. To make
sense of the above expressions, $C_1$ and $C_2$ can't be zero
simultaneously.

Substituting the above expressions into eqs. (\ref{Yeq}) and (\ref{newSeq}),
we get
\begin{equation}
Y^{\prime\prime} - \frac{\Delta}{2}\, (Y^\prime)^2 + Q(r)Y^\prime = R(r)
\label{NDf}
\end{equation}
and
\begin{equation}
  S^2=\Delta\left( Y^\prime
  -\frac{ 2C_2+\frac{1}{\Delta}( a C_3 +  \frac{\tilde{d}\, C_4}{D-2})
          r^{\tilde{d}} }
  { r(C_2+C_1r^{2\tilde{d}})} \right)^2
  +\frac{K}{(C_2+C_1r^{2\tilde{d}})^2 }r^{2\tilde{d}-2},
\label{Srslt}
\end{equation}
where
\begin{eqnarray}
    Q(r) & = & \frac{\tilde{d}+1}{r}
    + \frac{ 2C_2(\Delta-\tilde{d}) + ( a \, C_3 +  \frac{\tilde{d}\, C_4}
	{D-2}) r^{\tilde{d}} } {r(C_2 + C_1 r^{2\tilde{d}}) },
\label{Qdef}\\
R(r)&=&  { 1\over { r^2(C_2 +C_1 r^{2\tilde{d}})^2 } } 
\times  \left[ 2C_2^2(\Delta-\tilde{d})+2C_2 \left( a \, C_3 +  
\frac{\tilde{d}\, C_4}{D-2}\right)\, r^{\tilde{d}}  \right.
\nonumber \\
&  & \qquad \left.  +\left( 2C_1C_2
\tilde{d}(2\tilde{d}+1)  +\frac{1}{2}C_3^2 + { (D-3)\over 4(D-2) }\, C_4^2 
\right)\,  r^{2\tilde{d}} \right] , 
\label{Rdef}\\
K& = & C_3^2 - {1\over \Delta}  \, 
\left( a \, C_3 +  \frac{\tilde{d}\, C_4}{D-2}\right)^2 
\nonumber \\
& & \qquad + 
8 \tilde{d}(\tilde{d}+1)\, C_1 C_2  + {(D-3)\over 2 (D-2)}\, C_4^2.
\label{constK}
\end{eqnarray}

Notice that our system of equations of motion is an over determined system:
five unknown functions satisfying six equations. We have solved four
equations and there are two equations, eqs. (\ref{NDf}) and (\ref{Srslt}),
remaining with one unknown function $Y(r)$. These two equations have the
same form as the equations derived in \cite{ZhouZhu}. So the same proof given
 in 
\cite{ZhouZhu} can be used here to show that
these two equations actually give no constraints on $Y(r)$ and effectively there
is only one equation, i.e., we can solve either one of them and the other
one will
be satisfied automatically. In next section we will solve eq. (\ref{NDf}) 
completely.

\section{A Special Solution}

Now we start to solve eq. (\ref{NDf}). Let
\begin{equation}
g = \int \dd \, r \, e^{ \Delta Y - \int Q(r) dr },
\end{equation}
or,
\begin{equation}
 Y =  \frac{1}{\Delta}\left( \ln (g') + \int Q(r)dr \right),
\label{f2Y}
\end{equation}
eq. (\ref{NDf}) becomes
\begin{equation}
\left( \frac{g''}{g'} \right)' - { 1 \over 2 } \, \left( \frac{g''}{g'}
\right)^2 = \tilde{R}(r).
\label{lasty}
\end{equation}
Here
\begin{eqnarray}
  \tilde{R}(r) & = & \Delta R(r) - Q^\prime(r) - \frac{1}{2}\, Q^2(r)
  \nonumber \\
    & = & -\frac{\tilde{d}^2-1}{2\, r^2} +
    \frac{\tilde{\Lambda}\, r^{2\tilde{d}-2}}{ 2\,(C_2+C_1 r^{2\tilde{d}})^2 }
\label{NRdef}
\end{eqnarray}
with
\begin{equation}
  \tilde{\Lambda} =\Delta \, K - 4\, \tilde{d}^2\, C_1\, C_2
\end{equation}

The left-hand side of eq. (\ref{lasty}) is the well-known Schwarzian derivative
of the  funtion $g$ defined as:
\begin{equation}
S(g) \equiv  \left( \frac{g''}{g'} \right)' - \frac{1}{2} \left( \frac{g''}{g'}
  \right)^2.
\label{Schw}
\end{equation}
Thus, in order to solve eq. (\ref{lasty}), one must find a function $g$
such that
\begin{equation}
S(g)= -\frac{\tilde{d}^2-1}{2r^2} + \frac{\tilde{\Lambda}}{2}\
  \frac{ r^{ 2\tilde{d}-2 } }{ (C_2 + C_1 r^{2\tilde{d}})^2 }.
\label{target}
\end{equation}

\subsection{Notations and Conventions}

Let $f$ be an arbitrary function of $r$ and $f'$ be its derivative. The notation
$f(r)$ always represents the value of the function $f$ at $r$. Similarly, 
$f'(r)$ is the value of the function $f'$ at $r$.

For two functions $f$ and $g$, $f \,g$ is their product\footnote{We
denote $f\,f$ by $f^2$, $f\, f\, f$ by $f^3$, and so on.},
and $\displaystyle{\frac{f}{g}}$ is their division, while
the composition of two functions $f$ and $g$ is denoted by $f\circ g$.
These functions are defined by
\begin{eqnarray}
   (f\, g)(r)  & = & f(r) g(r), \\
  \left( \frac{f}{g} \right)(r) & = &  \frac{ f(r) }{ g(r) }, \\
  (f\circ g)(r) & = &  f( g(r) ).
\end{eqnarray}
For product of three or more functions we assume that the composition of two 
functions
has a higher rank of precedence of associavity than the products of functions.
Nevertheless the symbol
$\displaystyle{\frac{f}{g}}$ is considered to be a pure entity and can't be
breaked. To understand these conventions we have the following examples:
\begin{eqnarray}
  f\, g\circ h & = & g\circ h\, f = f\, (g\circ h), \\
   \frac{f}{g}\circ h & = &  \left( \frac{f}{g} \right)\circ h.
\end{eqnarray}

If $k\in \mathbb{C}$ or $\mathbb{R}$, we  define a function $l_k$ as
the multiplication of its variable with the number $k$, i.e.,
$$
  l_k(r)=k\, r.
$$
For convinence we also consider $k$ itself as a constant function taking the
value  $k$:
$$
  k(r) = k.
$$
Other functions such as the power function $r^s$ with real number $s$
and the exponential function $e^r$ are denoted by $p_s$ and $\exp$ and 
other elementary functions are denoted by their standard mathematical 
symbols.

With these conventions, we can derive the following derivative rules:
\begin{equation}
\begin{array}{ll}
  l'_k=k, &  \qquad \qquad k'=0, \\
  p'_s=s\, p_{s-1}  & \qquad \qquad   \arctan'=\frac{1}{1+p_2}, \\
  \exp'=\exp, & \qquad \qquad  \ln'=p_{-1},   
\end{array}
\end{equation}
and the derivative rule for composition of function is
\begin{equation}
  (f\circ g)'=g'\ f'\circ g
\end{equation}

\subsection{Some Properties of the Schwarzian Derivative}

 Now we list some elementary properties of the Schwarzian derivative.

(1) If $f$ and $g$ are two functions, we have 
\begin{equation}
  S(f\circ g)=(g')^2\ S(f)\circ g + S(g).
\end{equation}

(2) If $f=\displaystyle{ \frac{l_a+b}{l_c+d}}$ for some numbers $a$, $b$, $c$
and $d$, namely, $f(r)=\displaystyle{ \frac{ar+b}{cr+d} }$, then
\begin{equation}
  S\left(\frac{l_a+b}{l_c+d}\right)=0.
\end{equation}
This is the well-know fact that the fractional linear transformation is a
global conformal transformation of the complex sphere.  By using this result
we have
\begin{equation}
  S\left( \frac{l_a+b}{l_c+d}\circ g \right) = S(g).
\end{equation}
That is to say, if $g$ is a special solution of the equation $S(f)=R$, 
the general solution will be
\begin{equation}
  f = \frac{l_a+b}{l_c +d}\circ g = \frac{ag+b}{cg+d}
\label{fraclinear}
\end{equation}
with constants $a$, $b$, $c$, $d$ such that $ad-bc=1$.

(3) For $s\in\mathbb{R}$,  we have 
\begin{eqnarray}
 S(l_k)  & =&  0, \\
 S(p_s) & =&  -\frac{s^2-1}{2p_2}, \\
 S(\exp) & =&  - \frac{1}{2}, \\
  S(\ln) & =&  \frac{1}{2p_2}, \\
 S(\tan) & =&  2, \\
S(\arctan) & =&  - \frac{2}{(1+p_2)^2}.
\end{eqnarray}

\subsection{A Special Solution of Eq. (\ref{target})}

The function $\tilde{R}$ in Eq. (\ref{target}) is
\begin{equation}
  \tilde{R} = -\frac{\tilde{d}^2-1}{2p_2}
  + \frac{ \tilde{\Lambda} p_{2\tilde{d}-2} }{2}
  \,  \frac{1}{(C_2+C_1\, p_2)^2}\circ p_{\tilde{d}}.
\end{equation}
For $C_1 C_2 > 0$, we have 
\begin{eqnarray}
  \tilde{R} &=& S(p_{\tilde{d}}) + (p'_{\tilde{d}})^2\ \, 
S(h_1)\circ p_{\tilde{d}}
\nonumber \\
  &=& S(h_1\circ p_{\tilde{d}})
\label{Sh1}
\end{eqnarray}
where $h_1$ is a yet-unknown function such that
\begin{eqnarray}
  S(h_1) &=& \frac{\tilde{\Lambda}}{2\tilde{d}^2C_2^2}\
  \frac{1}{(1+\frac{C_1}{C_2}p_2)^2}
\nonumber \\
  &=& S(l_{\sqrt{C_1/C_2}}) + (l'_{\sqrt{C_1/C_2}})^2\ \frac{\tilde{\Lambda
}}
  {2\tilde{d}^2 C_1 C_2}\ \frac{1}{(1+p_2)^2}\circ l_{\sqrt{C_1/C_2}}
\nonumber \\
  &=& S(h_2\circ l_{\sqrt{C_1/C_2}}).
\label{Sh2}
\end{eqnarray}
Now we want to find a function $h_2$ such that
\begin{eqnarray}
  S(h_2) &=& \frac{\tilde{\Lambda}}{2\tilde{d}^2 C_1 C_2}\ \frac{1}{(1+
p_2)^2}
\nonumber \\
  &=& S(\arctan) + \frac{ \Delta \, K}{2\tilde{d}^2 C_1 C_2} \frac{1}{(1+
p_2)^2}
\nonumber \\
  &=& S(\arctan) + (\arctan')^2 \frac{\Delta \, K}{2\tilde{d}^2 C_1 C_2}\circ
 \arctan
\nonumber \\
  &=& S(h_3\circ \arctan)
\end{eqnarray}
where $K$ is the constant in eq. (\ref{constK}).
Note that in the third line we have considered $\displaystyle{\frac{\Delta \, K}
{2\,\tilde{d}^2 \,C_1\, C_2}}$ to be a constant function. The function $h_3$ 
in the
above is easy to find to be $\tan\circ l_k$ with
\begin{equation}
  k = \frac{1}{2\tilde{d}}\sqrt{ \frac{\Delta \, K}{C_1 C_2} },
\end{equation}
because
\begin{eqnarray}
  S(h_3) &=& \frac{\Delta \, K}{2\tilde{d}^2 C_1 C_2}
\nonumber \\
  &=& S(l_k) + (l'_k)^2\ S(\tan)\circ l_k
\nonumber \\
  &=& S(\tan\circ l_k).
\end{eqnarray}

Combining all the above steps, one finds that
\begin{equation}
  \tilde{R}=S(\tan\circ l_k\circ\arctan\circ l_{\sqrt{C_1/C_2}}\circ
  p_{\tilde{d}}).
\end{equation}
and a special solution of eq. (\ref{target}) is found to be:
\begin{equation}
  g_0=\tan\circ l_k\circ\arctan\circ l_{\sqrt{C_1/C_2}}\circ p_{\tilde{d}},
\end{equation}
namely,
\begin{equation}
  g_0(r)=\tan\left( k\arctan \sqrt{\frac{C_1}{C_2}}r^{\tilde{d}} \right).
\label{special}
\end{equation}
The special case considered in  \cite{ZhouZhu} corresponds to $k =1$.
By using Mathematica, we have checked that the above function is indeed
a solution of eq. (\ref{target}).

The general solution $f$ of eq. (\ref{target}) can be obtained according to
eq.~(\ref{fraclinear}). Here we have three independent constants. This is the
right number for a third order ordinary differential equation. So we obtain the
complete solution to eq. (\ref{target}). Substituting these  into
Eq.~(\ref{f2Y}), 
one can write down the complete solution to  eq.~(\ref{NDf}) which we turn in 
the next  section.

\section{The Complete Solution}

The general solution of eq. (\ref{target}) is obtained from the above special 
solution,
eq. (\ref{special}), by an  arbitrary $SL(2,R)$ transformation:
\begin{equation}
g(r) = { a_0 \, g_0(r) + b_0 \over c_0 \, g_0(r) + d_0 },
\end{equation}
where $a_0$, $b_0$, $c_0$ and $d_0$ consist of an $SL(2,R)$ matrix:
\begin{equation}
a_0 \, d_0 - b_0 \, c_0 = 1.
\end{equation}

With this general solution in hand one can check that the other equation 
(\ref{Srslt})
is also satisfied. Setting 
\begin{equation}
h(r) =  \arctan \sqrt{C_1\over C_2} \, r^{\tilde{d}} , 
\end{equation}
the complete solutions are:
\begin{eqnarray}
 \xi(r)  & = & \ln\left| C_1 + C_2 \, r^{-2 \tilde{d}}\right|,
\\
 \eta(r) & = & C_5 - {a \over \tilde{d}} \, \xi(r)  
+ {C_3\over \tilde{d}\sqrt{C_1 C_2}} \, h(r) ,
\\
 \beta(r) & = &  C_6 +  \xi(r)  + {C_4  \over \tilde{d}\sqrt{C_1 C_2}}\, h(r), 
\\
Y(r) & = &   - {2\over \Delta }\, \ln \left| C_7 \,\cos( k \, h(r) ) + C_8 \, 
\sin( k\, h(r) ) \right| 
 \nonumber \\
&  &  
- { 1\over \tilde{d}} \,  \xi(r)  
+ {a\, C_3  + { \tilde{d} \, C_4\over D-2} \over \tilde{d} \Delta \sqrt{
C_1 C_2}} \, h(r) ,
\end{eqnarray}
and
\begin{eqnarray}
f(r)  & = & \exp\left[ -C_6 - { C_4  \over \tilde{d} \, 
\sqrt{ C_1 \, C_2}}\, h(r) \right], 
\\
 A(r) & = &  {C_6 \over 2(D-2)}  + 
{ a \, C_3  + \left( 1 + {a^2 \over 2 \, \tilde{d} }\right) \, C_4  
\over (D-2) \Delta\sqrt{C_1 \, C_2} } \, h(r)   
\nonumber \\
& &  -  { 2 \tilde{d} \over \Delta \, (D-2)
} \,\ln \left| C_7 \,\cos( k \, h(r) ) + C_8 \, 
\sin( k\, h(r) ) \right|,
\\
 B(r) & = &  {C_6 \over2 ( D-2) }+ {1 \over \tilde{d} }\,   \xi(r)  
+ {a\, \left(   - 2 \, d \, C_3  + {a  } \, C_4 \right) 
\over 2\,  \tilde{d} \, \Delta\, (D-2  )\,\sqrt{C_1C_2} } \, h(r) 
\nonumber \\
& &  +  { 2 \, d \over \Delta\, (D-2) }  \, 
\ln \left| C_7 \,\cos( k \, h(r) ) + C_8 \, 
\sin( k\, h(r) ) \right| ,
 \\
 \phi(r)  & =&  C_5  + { 2 \, d \, C_3  - a \, C_4 \over 
(D-2) \Delta \sqrt{C_1\, C_2 } } \, h(r) 
\nonumber \\
& &  \qquad  +  { 2 \, a  \over \Delta }  \, 
\ln \left| C_7 \,\cos( k \, h(r) ) + C_8 \, 
\sin( k\, h(r) ) \right| , 
\\
 e^{\Lambda(r)} & = &  C_0 \, e^{ -\left( a \,C_5 + { \tilde{d} \, C_6 \over 
D-2 }\right) } \, 
\left[  C_9 + { 2 \over \sqrt{\Delta \, K} }
 \right. 
\nonumber \\
& &  \qquad \times \left. 
{ \sin(  k \, h(r)  ) \over 
C_7\, (C_7 \,\cos( k \, h(r) ) + C_8 \, 
\sin( k\, h(r) )  ) } \right], 
\end{eqnarray}
with
\begin{equation}
C_0^2\,  e^{ - a\, C_5 -  {\tilde{d} \, C_6 \over  D-2} }   = 
K \, (C_7^2 + C_8^2). 
\end{equation}
From the above explicit result we see that there are 9 constants of
integration. Most of them can be remove by chosing appropriate space-time
transformations. In the next section we will consider the the case for
$C_1 \, C_2 <0$ and requiring that the solution goes to flat space-time 
for $r \to \infty$. 

\section{ The case for $C_1 \, C_2 <0$. }

In this case the function $h(r)$ is changed to  
\begin{equation}
h(r)   =  \ln\left[  1 - \left( r_0 \over r \right)^{\tilde{d}} \over 
1 + \left( r_0 \over r \right)^{\tilde{d}}\right], 
\end{equation}
by omitting an overall constant and setting $C_2 = - r_0^{2 \, \tilde{d}}$,
($C_1 =1$). The complete solution is 
\begin{eqnarray}
 \xi(r) & = & \ln\left[ 1 - \left( r_0 \over r \right)^{2 \, \tilde{d}}
 \right] , 
\\
\eta(r) & = &- {  a \over \tilde{d} } \,  \xi(r)  + c_1 \, h(r) \, 
\\
\beta(r) & = & \xi(r) + c_2 \, h(r),
\\
Y(r) & = & - {  1 \over \tilde{d} } \,  \xi(r) + {1\over \Delta } \, 
\left( a \, c_1 + { \tilde{d}\, c_2 \over D-2} \right) \, h(r)
\nonumber \\
& & - {2 \over \Delta } \, \ln\left[ \cosh( \tilde{k} \, 
h(r) )  + c_3\, \sinh( \tilde{k} \, h(r) ) \right],
\end{eqnarray}
where  $c_i$'s are free constants and 
\begin{eqnarray}
\tilde{k}^2  &  = & - {\Delta \over 4 } \, \tilde{K},
\label{unknown}
\\
\tilde{K} & = & 
 c_1^2 - { 1\over \Delta}\,  \left( a \, c_1 + { \tilde{d}\, c_2 \over D-2}
 \right)^2
 - {2 (\tilde{d} +1) \over \tilde{d} } + { (D-3)\, c_2^2\over 2 (D-2) }  , 
\end{eqnarray}
where $\tilde{K}$ is assumed to be negative and 
\begin{eqnarray}
 f(r) & = &  \left[  1 - \left( r_0 \over r \right)^{\tilde{d}} \over 
1 + \left( r_0 \over r \right)^{\tilde{d}}\right]^{-c_2}
\\
 A(r) & = &   { \tilde{d}\, \left(  a \, c_1
 + \left( 1 + {a^2 \over 2 \, \tilde{d} }
\right) \,
c_2 \right) \over  \Delta (D-2)  } \, h(r) 
\nonumber \\
& & 
 - { 2 \, \tilde{d} \over \Delta (D-2) } 
\,  \ln\left[ \cosh( \tilde{k} \, h(r) ) 
+ c_3\, \sinh( \tilde{k} \, h(r) ) \right],
\\
B(r) & = &    { 1\over \tilde{d} } \, \xi(r) 
- { a\, (2 \, d \, c_1 - a \, c_2)  \over 2 \, \Delta \, {(D-2)} } \, h(r) 
\nonumber \\
& &  + { 2 \, {d} \over \Delta\, (D-2) } \, 
\ln\left[ \cosh( \tilde{k} \, h(r) ) 
+ c_3\, \sinh( \tilde{k} \, h(r) ) \right],
\\
\phi(r) & = &    {\tilde{d}\, (2 \, d \, c_1 - a \, c_2 ) \over \Delta 
\, (D-2)  } \,  h(r)   
\nonumber \\
& & + { 2 \, a \over \Delta} \, \ln\left[ \cosh( \tilde{k} \, h(r) ) 
+ c_3\, \sinh( \tilde{k} \, h(r) ) \right] , 
\\
e^{\Lambda(r) } & = &   c_0 \,  {\sinh( \tilde{k} \, h(r) ) 
\over  \cosh( \tilde{k} \,  h(r) )  + c_3\, \sinh( \tilde{k} \, h(r) ) }
+ \hbox{const.},
\end{eqnarray}
and 
\begin{equation}
c_0^2 = { 4 \over \Delta} \, ( c_3^2-1) .
\label{lastc0}
\end{equation}

For $\tilde{K} > 0$, we set $\displaystyle \tilde{k}^2 =   {\Delta \over 4 } \, 
\tilde{K}$, and the solution is  obtained by the simple substitution: $\cosh
\to \cos$ and $ \sinh \to \sin$. The last equation (\ref{lastc0}) is changed
to
\begin{equation}
c_0^2 = { 4 \over \Delta} \, ( c_3^2+1 ) .
\label{lastc00}
\end{equation}

The black brane solution found in \cite{Strominger, Duffa, Duffb} is a special 
case of the
above solution with $\displaystyle c_1 = { a\over \tilde{d}}$ and
$\displaystyle{ c_2 = -2 }$. The $\tilde{k}$ can be computed from
(\ref{unknown}) and we get $k =1$.
To make contact with the previous result, we should  change 
the isotropic coordinates used in this paper 
to the Schwarzschild-type coordinates. In Schwarzschild-type 
coordinate, the black brane solution obtained in  \cite{Strominger, Duffa,
Duffb} is
\begin{eqnarray}
\dd s^2 & = & f_-^{ 4 \tilde{d} \over \Delta\, (D-2) } \, \left[ 
-{ f_+ \over f_-} \, \dd t^2  + \dd x^i \dd x^i \right]  
\nonumber \\
& &  + 
f_-^{ 2 \, a^2 \over \Delta \, \tilde{d} } \, \left[ 
 {1\over f_+ \, f_-} \, \dd \tilde{r}^2 + \tilde{r}^2 \, \dd \Omega_{\tilde{d} 
+1}^2 \right],
\label{black}
\\
\phi & = &  - { 2 \, a \over \Delta } \, \ln f_-, 
\qquad f_{\pm} = 1 - \left( r_{\pm} \over \tilde{r} \right)^{\tilde{d}}. 
\end{eqnarray}
The transformation from  the isotropic coordinate $r$ to the 
 Schwarzschild-type coordinate $\tilde{r}$ is 
\begin{eqnarray}
 r  & = & \tilde{r} \left(  \sqrt{f_+} + \sqrt{f_-} \over 2\right)^{ 2 \over 
\tilde{d}} ,
\\
{\dd r^2 \over r^2} & = & {\dd \tilde{r}^2 \over \tilde{r}^2 } \, 
{1\over f_+ \, f_-}, 
\end{eqnarray}
and we have 
\begin{eqnarray}
& &\ln \left[ 1 - \left( r_0\over r\right)^{\tilde{d}} \right] = {1\over 2 }\, 
\ln f_+ + \ln \left[ 2\over \sqrt{f_+} + \sqrt{f_-} \right],
\\
& &\ln \left[ 1 +  \left( r_0\over r\right)^{\tilde{d}} \right] = {1\over 2 }\,
\ln f_- + \ln \left[ 2\over \sqrt{f_+} + \sqrt{f_-} \right],
\\
& & \ln (\cosh h(r) + c_3 \, \sinh h(r) ) = -{1\over 2} \,
( \ln f_+ + \ln f_-),
\end{eqnarray}
with 
\begin{eqnarray}  
r_0^{\tilde{d} } & = & {1\over 4} \, \left( r_+^{\tilde{d}} - r_-^{\tilde{d}}
 \right),
\\
c_3 & = & - { r_+^{\tilde{d}} + r_-^{\tilde{d} } \over 
 r_+^{\tilde{d}} - r_-^{\tilde{d} } }.
\end{eqnarray}
 We have then
\begin{eqnarray}
f(r) & = & { f_+ \over f_-}, 
\\ 
A(r) & = & - {2 \, \tilde{d} \over \Delta \, (D-2) } \, \left( h(r) + 
\ln (\cosh h(r) + c_3 \, \sinh h(r) )  \right)
\nonumber \\
& = & { 2 \, \tilde{d} \over \Delta \, (D-2) }  \, \ln f_-, 
\\
B(r) & = & {1\over \tilde{d}} \, \xi(r) - { a^2 \over \Delta \, \tilde{d} } \, 
h(r) + { 2 \, d\over \Delta \, (D-2) } \, 
\ln (\cosh h(r) + c_3 \, \sinh h(r) )  
\nonumber \\
 & = &  {2 \over \tilde{d} } \, \ln { 2 \over \sqrt{f_+} + \sqrt{f_-}} 
+ { a^2 \over \Delta \, \tilde{d} } \, \ln f_-,
\\
\phi(r) & = & {2 \, a\over \Delta} \, \left( h(r) + 
\ln (\cosh h(r) + c_3 \, \sinh h(r) )  \right)
= - {2 \, a\over \Delta} \, \ln f_-.
\end{eqnarray}
The above  results gave  exactly the black brane solution (\ref{black}). 

From the above results we see that the black brane solution is a two-parameter
solution. Our complete solution is a four-parameter solution. A detail study of 
the physical property of this more general  solution is left for a future
publication.

\section *{Acknowledgments}
We would like to thank Han-Ying Guo, Yi-hong Gao, Ke Wu, Ming Yu,
Zhu-jun Zheng and Zhong-Yuan Zhu for discussions. This work is supported in
part by funds from Chinese National Science Fundation and Pandeng Project.

\end{document}